\documentclass{amsbook}
\usepackage{amsmath}
\usepackage{amssymb}
\usepackage{dsfont}
\usepackage{multicol}
\usepackage{proof}
\usepackage{setspace}
\usepackage{booktabs}
\usepackage[dvipsnames]{xcolor}
\usepackage{stmaryrd}
\usepackage{mathtools}

\theoremstyle{definition}

\theoremstyle{remark}

\numberwithin{equation}{section}

\def\InputModeColorName{MidnightBlue}
\def\OutputModeColorName{Maroon}
\newcommand\InputMode[1]{{\color{\InputModeColorName}{#1}}}
\newcommand\OutputMode[1]{{\color{\OutputModeColorName}{#1}}}

\newcommand\isprop[1]{\ensuremath{\InputMode{#1}\;\mathit{prop}}}
\newcommand\isset[1]{\ensuremath{\InputMode{#1}\;\mathit{type}}}
\newcommand\eisset[2]{\ensuremath{\InputMode{#1}=\InputMode{#2}\;\mathit{type}}}
\newcommand\istrue[1]{\ensuremath{\InputMode{#1}\;\mathit{true}}}
\newcommand\isctx[1]{\ensuremath{\InputMode{#1}\;\mathit{ctx}}}
\newcommand\fresh[2]{\ensuremath{\InputMode{#1}\mathrel{\#}\InputMode{#2}}}
\newcommand\sequent[2]{\ensuremath{\InputMode{#1}\gg#2}}
\newcommand\hyp[2]{\ensuremath{#1\ (#2)}}

\newcommand\genj[2]{\ensuremath{\lvert_{\InputMode{#1}}\,#2}}
\newcommand\ver[2]{\ensuremath{\InputMode{#1}\in\InputMode{#2}}}
\newcommand\isenv[2]{\ensuremath{\InputMode{#1}\in^\star\InputMode{#2}}}
\newcommand\eqenv[3]{\ensuremath{\InputMode{#1}=\InputMode{#2}\in^\star\InputMode{#3}}}
\newcommand\ever[3]{\ensuremath{\InputMode{#1}=\InputMode{#2}\in\InputMode{#3}}}
\newcommand\ofty[2]{\ensuremath{\InputMode{#1}:\InputMode{#2}}}
\newcommand\eofty[3]{\ensuremath{\InputMode{#1}=\InputMode{#2}:\InputMode{#3}}}
\newcommand\reduce[2]{\ensuremath{\InputMode{#1}\Downarrow\OutputMode{#2}}}
\newcommand\Match[2]{\ensuremath{\InputMode{#1}\equiv\OutputMode{#2}}}

\newcommand\naturals{\ensuremath{\mathbb{N}}}

\newcommand\lfsequent[2]{\ensuremath{\InputMode{#1}\vdash#2}}
\newcommand\prf[1]{\ensuremath{\mathsf{Prf}(#1)}}
\newcommand\MLLF{\textbf{MLLF}}

\newcommand\type[1]{\ensuremath{\InputMode{#1}:\mathit{type}}}
\newcommand\etype[2]{\ensuremath{\InputMode{#1}=\InputMode{#2}:\mathit{type}}}

\newcommand\product[2]{\ensuremath{#1\&#2}}
\newcommand\pair[2]{\ensuremath{\langle #1,#2\rangle}}
\newcommand\fst[1]{\ensuremath{\mathsf{fst}(#1)}}
\newcommand\snd[1]{\ensuremath{\mathsf{snd}(#1)}}
\newcommand\abort[2]{\ensuremath{\mathsf{abort}(#1;#2)}}

\newcommand\tyunit{\ensuremath{\mathsf{unit}}}
\newcommand\tyvoid{\ensuremath{\mathsf{void}}}
\newcommand\typrod[3]{\ensuremath{\mathsf{\Pi}(#1;#2.#3)}}
\newcommand\lam[2]{\ensuremath{\mathsf{\lambda}(#1.#2)}}
\newcommand\ap[2]{\ensuremath{\mathsf{ap}(#1;#2)}}

\begin{document}
\title{Type theory and its meaning explanations}
\author{Jonathan Sterling}

\thanks{Thanks to Bob Harper, Peter Dybjer, Bengt Nordstr\"om, Carlo Angiuli
and Darin Morrison for invaluable conversations which helped sharpen my view of
the meaning explanations for computational and intensional type theories. All
mistakes are mine.}

\maketitle

\begin{abstract}
  At the heart of intuitionistic type theory lies an intuitive semantics called
  the ``meaning explanations''; crucially, when meaning explanations are taken as
  definitive for type theory, the core notion is no longer ``proof'' but
  ``verification''. We'll explore how type theories of this sort arise naturally
  as enrichments of logical theories with further judgements, and contrast this
  with modern proof-theoretic type theories which interpret the judgements and
  proofs of logics, not their propositions and verifications.
\end{abstract}

\onehalfspacing

\tableofcontents

\chapter{Logical Theories}

To start, we will consider the notion of a \emph{logical theory}; in
my mind, it begins with a species  of judgements that can be
proposed, asserted, and (if they are evident) known.

\section{Judgements of a logical theory}

The basic forms of judgement for a logical theory will be
\framebox{$\isprop{P}$} and \framebox{$\istrue{P}$}; and what is $P$?  It is a
member of the species of terms, which are made meaningful in the course of
making the judgement $\isprop{P}$ (``$P$ is a proposition'') evident for a
proposition $P$.

The forms of judgement may be construed as containing
\InputMode{\textit{inputs}} and \OutputMode{\textit{outputs}}; an
\InputMode{\textit{input}} is something which is inspected in the course of
knowing a judgement, whereas an \OutputMode{\textit{output}} is something which
is synthesized (or created) in the course of knowing a judgement. The positions
of \InputMode{\textit{inputs}} and \OutputMode{\textit{outputs}} in a judgement
constitute what is called its \emph{mode}, and we color-code it in this
presentation for clarity.\footnote{We will not see any judgements with
  \OutputMode{\textit{outputs}} at first, but it will become necessary as soon
  as we consider judgements about computation, where the reduction of a term is
synthesized from the redex. Modes may be used to construe a judgement as having
algorithmic content.}

To each judgement is assigned a \emph{meaning explanation}, which explicates
the knowledge-theoretic content of the judgement. For a judgement
$\mathcal{J}$, a meaning explanation should be in the form:

\begin{quote}
  To know $\mathcal{J}$ is to know...
\end{quote}

The meaning of the judgement $\isprop{P}$ is, then, as follows:

\begin{quote}
  To know $\isprop{P}$ is to know that $P$ is a proposition, which is to know
  what would count as a direct verification of $P$.
\end{quote}

So if a symbol $P$ is taken to denote a proposition, we must know
\emph{what sort of thing} is to be taken as a direct verification of
$P$, and this is understood as part of the definition of $P$. A
``direct verification'' is understood in constrast with an ``indirect
verification'', which is to be thought of as a means or plan for
verifying the proposition; these distinctions will be explained in
more detail later on. Now, the judgement $\istrue{P}$ (``$P$ is
true'') is only meaningful in case we know $\isprop{P}$ (this is
called a presupposition). Then the meaning of $\istrue{P}$ is as
follows:

\begin{quote}
  To know $\istrue{P}$ is to have a verification of $P$.
\end{quote}

From the (implicit) presupposition $\isprop{P}$, we already know what
counts as a verification, so the meaning explanation is
well-defined. Note that having a means or plan for verifying $P$ is
equivalent to having a (direct) verification; this follows from the
fact that one may put into action a plan for verifying $P$ and achieve
such a verification, and likewise, it is possible to propound a plan
of verification by appeal to an existing verification.

\section{Higher-order judgements}

The judgements we have described so far are ``categorical'' in the sense that
they are made without assumption or generality.

\subsection{Hypothetical judgement}

We will need to define a further form of judgement, which is called
``hypothetical'', and this is the judgement under hypothesis
\framebox{$\hyp{\mathcal{J}}{\mathcal{J}'}$}, pronounced
``$\mathcal{J}$ under the assumption $\mathcal{J}'$''.Its meaning
explanation is as follows:

\begin{quote}
  To know the judgement $\hyp{\mathcal{J}}{\mathcal{J}'}$ is to know
  the judgement $\mathcal{J}$ assuming you know the
  judgement $\mathcal{J}'$.
\end{quote}

Hypothetical judgement may be iterated, and
$\hyp{\mathcal{J}}{\mathcal{J}_1,\mathcal{J}_2}$ will be used as notation for
for $\hyp{\hyp{\mathcal{J}}{\mathcal{J}_2}}{\mathcal{J}_1}$.

\subsection{General judgement}

Another kind of higher order judgement is ``general judgement'', which
is judgement with respect to a variable,
\framebox{$\genj{x}{\mathcal{J}}$}, pronounced ``for an arbitrary $x$,
$\mathcal{J}$''. The meaning explanation for this new judgement is as
follows:

\begin{quote}
  To know the judgement $\genj{x}{\mathcal{J}}$ is, to know
$[E/x]\mathcal{J}$ (i.e. the substitution of $E$ for $x$ in the
expression $\mathcal{J}$) for any arbitrary expression $E$,
\footnote{Technically, $E$ is qualified as being of the same valence
as $x$, but because we have not developed a formal theory of
expressions in this presentation, I choose to ignore this issue.}
\end{quote}

As far as notation is concerned, the bar symbol binds the least
tightly of all the other notations we have considered. Likewise,
general judgement may be iterated, and the notation
$\genj{x,y}{\mathcal{J}}$ will be used as notation for
$\genj{x}{\genj{y}{\mathcal{J}}}$.

\subsection{Hypothetico-general judgement}

When hypothetical judgement is used inside general judgement, as in
$\genj{x}{\hyp{\istrue{A(x)}}{\istrue{B(x)}}}$, we term the whole
thing a ``hypothetico-general'' judgement. One thing bears clarifying,
which is, Why do we write $\hyp{\istrue{P}}{\istrue{P}}$ rather than
$\genj{P}{\hyp{\istrue{P}}{\istrue{P}}}$?

The former is really not a single judgement, but rather a
\emph{scheme} for judgements, where $P$ is intended to be replaced
with a concrete expression by the person asserting the judgements. On
the other hand, the latter is itself a single judgement which may be
asserted all on its own.

\section{Propositions and verifications}
Now that we have propounded and explained the minimal system of judgements for
a logical theory, let us populate it with propositions. First, we have falsity
$\bot$, and we wish to make $\isprop{\bot}$ evident; to do this, we simply state
what counts as a direct verification of $\bot$: there is no
direct verification of $\bot$.

The next basic proposition is trivial truth $\top$, and to make
$\isprop{\top}$ evident, we state that a direct verification of $\top$
is trivial. The definition of $\top$ thus validates the judgement
$\istrue{\top}$ (i.e. that we have a verification of $\top$; this
is immediate).

Next, let us define conjunction; in doing so, we will make evident the
hypothetical judgement
\[\hyp{\isprop{P\land Q}}{\isprop{P},\isprop{Q}}\]
Equivalently, we can display this as a rule of inference:\footnote{
Evident hypothetical judgements are often written as rules, i.e.
\[
  \infer{\textit{conclusion}}{\textit{premise}}
\]
rather than $\hyp{\textit{conclusion}}{\textit{premise}}$. It must be
stressed that only \emph{evident}/\emph{known} judgements may be
written in this way.
}
\[
  \infer{
    \isprop{P\land Q}
  }{
    \isprop{P} &
    \isprop{Q}
  }
\]

A direct verification of $P\land Q$ consists in a verification of $P$
and a verification of $Q$; this validates the assertion of the judgement
$\hyp{\istrue{P\land Q}}{\istrue{P},\istrue{Q}}$. Because it is a valid
inference, we can write it as an inference rule:
\[
  \infer{
    \istrue{P\land Q}
  }{
    \istrue{P} &
    \istrue{Q}
  }
\]

A direct verification of $P\lor Q$ may be got either from a verification
of $P$ or one of $Q$. From this definition we know $\hyp{\isprop{P\lor
Q}}{\isprop{P},\isprop{Q}}$, or
\[
  \infer{
    \isprop{P\lor Q}
  }{
    \isprop{P} &
    \isprop{Q}
  }
\]
The verification conditions of disjunction give
rise to two evident judgements $\hyp{\istrue{P\lor Q}}{\istrue{P}}$ and
$\hyp{\istrue{P\lor Q}}{\istrue{Q}}$, which we can write as inference rules:
\begin{gather*}
  \infer{
    \istrue{P\lor Q}
  }{
    \istrue{P}
  }
  \qquad
  \infer{
    \istrue{P\lor Q}
  }{
    \istrue{Q}
  }
\end{gather*}

Finally, we must define the circumstances under which $P\supset Q$ is
a proposition (i.e.\ when $\isprop{P\supset Q}$ is evident). And we
intend this to be under the circumstances that $P$ is a proposition,
and also that $Q$ is a proposition assuming that $P$ is true. In other
words, $\hyp{\isprop{P\supset Q}}{\isprop{P},
\hyp{\isprop{Q}}{\istrue{P}}}$, or
\[
  \infer{
    \isprop{P\supset Q}
  }{
    \isprop{P} &
    \hyp{\isprop{Q}}{\istrue{P}}
  }
\]

Now, to validate this judgement will be a bit more complicated than the
previous ones. But by unfolding the meaning explanations for hypothetical
judgement, proposition-hood and truth of a proposition, we arrive at the
following explanation:
\begin{quote}
  To know $\hyp{\isprop{P\supset Q}}{\isprop{P},\hyp{\isprop{Q}}{\istrue{P}}}$
  is to know what counts as a direct verification of $P\supset Q$ when one
  knows what counts as a direct verification of $P$, and, when one has such a
  verification, what counts as a direct verification of $Q$.\footnote{Note that
  unless $\istrue{P}$, it need not be evident that $\isprop{Q}$; in other
  words, $Q$ only has to be a proposition if $P$ is true. It would also be
  acceptable to give a stronger definition to implication, but this is the one
  accepted by Martin-L\"of.}
\end{quote}

If the judgement $\hyp{\isprop{P\supset
Q}}{\isprop{P},\hyp{\isprop{Q}}{\istrue{P}}}$ is going to be made evident, then
we must come up with what should count as a direct verification of $P\supset Q$
under the assumptions described above.

And so to have a direct verification of $P\supset Q$ is to have a
verification of $Q$ assuming that one has one of $P$; this is the
meaning of implication, and it validates the judgement
$\hyp{\istrue{P\supset Q}}{\hyp{\istrue{Q}}{\istrue{P}}}$, and may be
written as an inference rule as follows:
\[
  \infer{
    \istrue{P\supset Q}
  }{
    \hyp{\istrue{Q}}{\istrue{P}}
  }
\]

\section{Judgements for verifications}

So far, we have given judgements which define what it means to be a
proposition, namely $\isprop{P}$, and thence for each proposition, we
have by definition a notion of what should count as a verification of
that proposition. And we have a judgement $\istrue{P}$, which in its
assertion means that one has (a way to obtain) such a verification of
$P$, but we have not considered any judgements which actually refer to
the verifications themselves symbolically.

It is a hallmark of Martin-L\"of's program to resolve the
contradiction between syntax and semantics not by choosing symbols
over meanings or meanings over symbols, but by endowing symbols with
meaning in the course of knowing the evident judgements. \textbf{As
such, $P$ is a symbol, but when we assert $\isprop{P}$ we are saying
that we know what proposition $P$ denotes.}

A similar thing can be done with verifications themselves, by
representing them with symbols in the same way we have done for the
propositions. And then, we can consider a judgement such as ``$M$ is a
verification of $P$'', and in knowing that judgement, we know what
verification $M$ is meant to denote. In practice, this judgement has
been written in several ways:\\[10pt]
\begin{tabular}{c|l}
  Notation & Pronunciation \\ \hline
  $\InputMode{M}\in \InputMode{P}$ & $M$ is an element of $P$\\
  $\InputMode{M}\Vdash \InputMode{P}$ & $M$ realizes $P$\\
  $\InputMode{P}\ {\color{Gray}\llcorner\mathsf{ext}\;} \OutputMode{M}{\color{Gray}\lrcorner}$ & $P$ is witnessed by $M$
\end{tabular}\\

But they all mean the same thing, so we will choose the notation
\framebox{$\ver{M}{P}$} and pronounce it ``$M$ verifies $P$''. Tentatively, the
following defective meaning explanation could be given:
\begin{quote}
  * To know $\ver{M}{P}$ is to know that $M$ is a verification of $P$.
\end{quote}

But now that we have started to assign expressions to verifications,
we must be more careful about differentiating \emph{direct
verifications} (which we will call ``canonical'') from \emph{indirect
verifications} (which we will call ``non-canonical''). So the domain
of expressions must itself be accorded with a notion of reduction to
canonical form, and this corresponds with putting into action a plan
of verification in order to get a direct (canonical) verification;
reduction to canonical form will be represented by a judgement
\framebox{$\reduce{M}{M'}$}, pronounced ``$M$ evaluates to $M'$''.

\begin{quote}
  To know $\reduce{M}{M'}$ is to know that $M$ is an expression which reduces to
  a canonical form $M'$.
\end{quote}

An example of an evident reduction judgement in elementary mathematics
would be $\reduce{3+4}{7}$; note that $\reduce{3+4}{1+6}$ is, on the
other hand, not evident, since this judgement describes reduction to
\emph{canonical} form, whereas $1+6$ is not a canonical number.

Now, we can correct the previous meaning explanation as follows:

\begin{quote}
  To know $\ver{M}{P}$ is to know an $M'$ such that $\reduce{M}{M'}$ and $M'$ is a
  canonical (direct) verification of $P$.
\end{quote}

If it is not yet clear why it would have been a mistake to fail to use
the notion of reduction to canonical form in the above meaning
explanation, consider that each time a proposition is defined, it
should be possible to do so without knowing what other propositions
exist in the theory. But if we consider non-canonical forms (as would
be necessary if we omitted the $\reduce{M}{M'}$ premise), then we
would have to fix in advance all the possible non-canonical forms in
the computation system in the course of defining each proposition. As
such, the open-ended nature of the logic would be destroyed; in a
later chapter, the seriousness of this problem will be made even more
clear.

The meaning explanation for $\isprop{P}$ must be accordingly modified to take
into account the computational behavior of expressions:

\begin{quote}
  To know $\isprop{P}$ is to know a $P'$ such that $\reduce{P}{P'}$ and $P'$ is a
  canonical proposition, which is to say, that one knows what counts as a
  canonical verification for $P'$.
\end{quote}

In practice, when it is clear that $P$ is canonical, then we will simply say,
``To know $\isprop{P}$ is to know what counts as a canonical verification of
$P$''. As an example, then, we will update the evidence of the following assertion:
\[\hyp{\isprop{P\supset Q}}{\isprop{P},\hyp{\isprop{Q}}{\istrue{P}}}
\]
The meaning
of this, expanded into spoken language, is as follows:
\begin{quote}
  To know $\hyp{\isprop{P\supset
Q}}{\isprop{P},\hyp{\isprop{Q}}{\istrue{P}}}$ is to know what counts
  as a canonical (direct) verification of $P\supset Q$ under the
  circumstances that $\reduce{P}{P'}$, such that one knows what counts
  as a canonical verification $P'$, and, if one has such a verification,
  $\reduce{Q}{Q'}$ such that one knows what counts as a canonical
  verification of $Q'$.
\end{quote}

And the above judgement is evident, since we will say that a canonical
verification of $P\supset Q$ is an expression $\lambda x. E$ such that
we know the hypothetico-general judgement
$\genj{x}{\hyp{\ver{E}{Q}}{\ver{x}{P}}}$. This validates the assertion
$\hyp{\ver{\lambda x.E}{P\supset
Q}}{\genj{x}{\hyp{\ver{E}{Q}}{\ver{x}{P}}}}$, or, written as an
inference rule:

\[
  \infer{
    \ver{\lambda x.E}{P\supset Q}
  }{
    \genj{x}{\hyp{\ver{E}{Q}}{\ver{x}{P}}}
  }
\]

By the addition of this judgement, we have graduated from a logical theory to a
type theory, in the sense of \emph{Constructive Mathematics and Computer
Programming} (Martin-L\"of, 1979). In fact, we may dispense with the original
$\istrue{P}$ form of judgement by \emph{defining} it in terms of the new
$\ver{M}{P}$ judgement as follows:
\[
  \infer{
    \istrue{P}
  }{
    \ver{M}{P}
  }
\]


\chapter{Computational Type Theories}

As alluded to at the end of the previous chapter, we may add a judgement
\ver{M}{A} which deals directly with the objects $M$ which verify the
propositions (or types) $P$. We will develop a full theory of dependent types
in the sense of Martin-L\"of.\footnote{Please note that the judgements given here, and
their meaning explanations, are \emph{not} the same as those used in Constable
et al's ``Computational Type Theory'' and Nuprl. In this chapter, we use the
term ``computational type theory'' in a general sense to characterize a family
of type theories which have their origin in Martin-L\"of's 1979 paper
\emph{Constructive Mathematics and Computer Programming}.}

\section{The categorical judgements}

Because we will need to consider the introduction of types which do not have a
trivial (intensional) equality relation, we must first amend the meaning
explanations for some of our judgements, and add a few new forms of judgement.
First, we will refer to \emph{types} rather than \emph{propositions} in order
to emphasize the generality of the theory; in some presentations, the word
\emph{set} is used instead.

The meaning of hypothetical and general judgement are the same as in
the previous chapter, and so we will not reproduce them here. The
first form of judgement is \framebox{\isset{A}}, and its meaning
explanation is as follows:

\begin{quote}
  To know $\isset{A}$ is to know an $A'$ such that $\reduce{A}{A'}$,
  and you know what counts as a canonical verification of $A'$ and when
  two such verifications are equal.
\end{quote}

The next form of judgement is \framebox{\ver{M}{A}}, which remains
the same as before:

\begin{quote}
  To know $\ver{M}{A}$ (presupposing $\isset{A}$) is to know an $M'$ such that
  $\reduce{M}{M'}$ and $M'$ is a canonical verification of $A$.
\end{quote}

We'll need to add new judgements for equality (equality of
verifications, and equality of types respectively). First, equality of
verifications is written \framebox{$\ever{M}{N}{A}$}, and means the
following:

\begin{quote}
  To know $\ever{M}{N}{A}$ (presupposing $\ver{M}{A}$ and $\ver{N}{A}$,
  and thence $\isset{A}$) is to know that the values of $M$ and $N$ are
  equal as canonical verifications of $A$.
\end{quote}

The fact that $M$ and $N$ reduce to canonical values which verify $A$
is known from the presuppositions of the judgement; and what it means
for them to be equal as such is known from the evidence of the
presupposition $\isset{A}$ which is obtained from the other
presuppositions.

For equality of types \framebox{$\eisset{A}{B}$}, there are a number of
possible meaning explanations, but we'll use the one that Martin-L\"of
used starting in 1979:

\begin{quote}
  To know $\eisset{A}{B}$ (presupposing $\isset{A}$ and $\isset{B}$) is to know
$\genj{M}{\hyp{\ver{M}{B}}{\ver{M}{A}}}$ and
$\genj{M}{\hyp{\ver{M}{A}}{\ver{M}{B}}}$, and moreover
$\genj{M,N}{\hyp{\ever{M}{N}{B}}{\ever{M}{N}{A}}}$ and
$\genj{M,N}{\hyp{\ever{M}{N}{A}}{\ever{M}{N}{B}}}$.
\end{quote}

In other words, two types are equal when they have the same canonical
verifications, and moreover, the same equality relation over their canonical
verifications. Note that there are other possible explanations for type
equality, including more intensional ones that appeal to the syntactic
structure of type expressions, and these turn out to be more useful in proof
assistants for practical reasons. However, the extensional equality that we
have expounded is the easiest and most obvious one to formulate, so we will
use it here.

\section{The functional sequent judgement}

Now, because we are allowing the definition of types with arbitrary equivalence
relations, we cannot use plain hypothetico-general judgement in the course of
defining our types. For instance, if we were going to try and define the
function type $A\supset B$ in the same way as we did in the previous chapter,
we would permit ``functions'' which are not in fact functional, i.e.\ they do
not take equal inputs to equal outputs. As such, we will need to bake
functionality (also called extensionality) into the definition of functions,
and since we will need this in many other places, we elect to simplify our
definitions by baking it into a single judgement which is meant to be used
instead of plain hypothetico-general judgement.

The judgement which expresses simultaneously generality, hypothesis and
functionality has been written in multiple ways. Martin-L\"of has always
written it as $\hyp{\mathcal{J}}{\Gamma}$, but this is a confusing notation
because it appears as though it is merely a hypothetical judgement (but it is
much more, as will be seen). Very frequently, it is written with a turnstile,
$\Gamma\vdash\mathcal{J}$, and in the early literature surrounding Constable's
Computational Type Theory and Nuprl, it was written
\framebox{$\sequent{\Gamma}{\mathcal{J}}$}; we choose this last option to avoid
confusion with a similar judgement form which appears in proof-theoretic,
intensional type theories; we'll call the judgement form a ``(functional) sequent''.

We will define several forms of judgment simultaneously with the sequent judgments:
\framebox{$\isctx{\Gamma}$} (``$\Gamma$ is a context''),
\framebox{$\fresh{x}{\Gamma}$} (``$x$ is fresh in $\Gamma$'').

\begin{quote}
  To know $\isctx\Gamma$\ is to know that $\Match{\Gamma}{\cdot}$, or it is to know
  a variable $x$ and expressions $\Delta,A$ such that $\Match{\Gamma}{\Delta, x:A}$
  and $\sequent{\Delta}{\isset{A}}$, and $\fresh{x}{\Delta}$.
\end{quote}

\begin{quote}
  To know $\fresh{x}{\Gamma}$\ (presupposing $\isctx\Gamma$) is to know that
  $\Match{\Gamma}{\cdot}$, or, if $\Match{\Gamma}{\Delta,y:A}$
  such that $x$ is not $y$ and $\fresh{x}{\Delta}$.
\end{quote}

In other words, the well-formed contexts are inductively generated by the
following grammar:
\begin{gather*}
  \infer{\isctx\cdot}{}
  \qquad
  \infer{
    \isctx{\Gamma,x:A}
  }{
    \isctx\Gamma &
    \sequent{\Gamma}{\isset{A}} &
    \fresh{x}{\Gamma}
  }\\
  \infer{\fresh{x}{\cdot}}{}
  \qquad
  \infer{
    \fresh{x}{\Gamma,y:A}
  }{
    \fresh{x}{\Gamma}
  }
\end{gather*}

Next, we define the notion of an \emph{environment} or \emph{substitution} for
a context, $\framebox{\isenv{\rho}{\Gamma}}$. It is possible to define this
more generally as functional maps from one context to another, but for our
purposes, this is the only case we need.
\begin{quote}
  To know $\isenv{\rho}{\Gamma}$, presupposing $\isctx{\Gamma}$, is to know, if
  $\Match{\Gamma}{\cdot}$, then $\Match{\rho}{\cdot}$; otherwise, if
  $\Match{\Gamma}{\Delta, x:A}$ then $\Match{\rho}{\rho', M}$ such that
  $\isenv{\rho'}{\Delta}$ and $\ver{[\rho'/\Delta]M}{[\rho'/\Delta]A}$. The substitution
  $[\rho/\Gamma]$ is effected by replacing each variable in $\Gamma$ with the
  corresponding term in $\rho$.
\end{quote}

We will also specify when two environments are equal,
$\framebox{\eqenv{\rho_0}{\rho_1}{\Gamma}}$:
\begin{quote}
  To know $\eqenv{\rho_0}{\rho_1}{\Gamma}$, presupposing $\isctx{\Gamma}$,
  $\isenv{\rho_0}{\Gamma}$ and $\isenv{\rho_1}{\Gamma}$ is to know, if
  $\Match{\Gamma}{\cdot}$, then $\Match{\rho_0}{\cdot}$ and
  $\Match{\rho_1}{\cdot}$; otherwise, if $\Match{\Gamma}{\Delta,x:A}$ and
  thence $\Match{\rho_0}{\rho_0',M_0}$ and $\Match{\rho_1}{\rho_1',M_1}$, then
  $\eqenv{\rho_0'}{\rho_1'}{\Delta}$ and
  $\ever{[\rho_0/\Delta]M_0}{[\rho_1/\Delta]M_1}{[\rho_0/\Delta]A}$.
\end{quote}

We will say that $\sequent{\Gamma}{\mathcal{J}}$ is only a judgement under the
presuppositions that $\isctx\Gamma$\ and that $\mathcal{J}$ is a categorical
judgement of the form $\isset{A}$, $\eisset{A}{B}$, $\ver{M}{A}$, or
$\ever{M}{N}{A}$.  Now we may begin giving the meaning explanations for
$\sequent{\Gamma}{\mathcal{J}}$, starting with typehood with respect to a
context:

\begin{quote}
  To know $\sequent{\Gamma}{\isset{A}}$ (presupposing $\isctx\Gamma$) to know
  $
    \genj{\rho}{
      \hyp{
        \isset{[\rho/\Gamma]A}
      }{
        \isenv{\rho}{\Gamma}
      }
    }
  $
  and moreover, to know
  $
    \genj{\rho_0,\rho_1}{
      \hyp{
        \eisset{[\rho_0/\Gamma]A}{[\rho_1/\Gamma]A}
      }{
        \eqenv{\rho_0}{\rho_1}{\Gamma}
      }
    }
  $.
\end{quote}

We can explain type equality sequents in a similar way:

\begin{quote}

  To know $\sequent{\Gamma}{\eisset{A}{B}}$ (presupposing $\isctx\Gamma$,
  $\sequent{\Gamma}{\isset{A}}$, $\sequent{\Gamma}{\isset{B}}$) is to know
  \[
    \genj{\rho}{
      \hyp{
        \eisset{[\rho/\Gamma]A}{[\rho/\Gamma]B}
      }{
        \isenv{\rho}{\Gamma}
      }
    }
  \]
  and moreover, to know
  \[
    \genj{\rho_0,\rho_1}{
      \hyp{
        \eisset{[\rho_0/\Gamma]A}{[\rho_1/\Gamma]B}
      }{
        \eqenv{\rho_0}{\rho_1}{\Gamma}
      }
    }
  \]
\end{quote}

Next, the meaning of membership sequents is explained:

\begin{quote}
  To know $\sequent{\Gamma}{\ver{M}{A}}$ (presupposing $\isctx\Gamma$, $\sequent{\Gamma}{\isset{A}}$) is to know
  $
    \genj{\rho}{
      \hyp{
        \ver{[\rho/\Gamma]M}{[\rho/\Gamma]A}
      }{
        \isenv{\rho}{\Gamma}
      }
    }
  $
  and moreover, to know
  \[
    \genj{\rho_0,\rho_1}{
      \hyp{
        \ever{[\rho_0/\Gamma]M}{[\rho_1/\Gamma]M}{[\rho_0/\Gamma]A}
      }{
        \eqenv{\rho_0}{\rho_1}{\Gamma}
      }
    }
  \]
\end{quote}

Finally, member equality sequents have an analogous explanation:

\begin{quote}
  To know $\sequent{\Gamma}{\ever{M}{M'}{A}}$ (presupposing $\isctx\Gamma$, $\sequent{\Gamma}{\isset{A}}$) is to know
  $
    \genj{\rho}{
      \hyp{
        \ever{[\rho/\Gamma]M}{[\rho/\Gamma]N}{[\rho/\Gamma]A}
      }{
        \isenv{\rho}{\Gamma}
      }
    }
  $
  and moreover, to know
  \[
    \genj{\rho_0,\rho_1}{
      \hyp{
        \ever{[\rho_0/\Gamma]M}{[\rho_1/\Gamma]N}{[\rho_0/\Gamma]A}
      }{
        \eqenv{\rho_0}{\rho_1}{\Gamma}
      }
    }
  \]
\end{quote}

The simultaneous definition of multiple judgements may seem at first
concerning, but it can be shown to be non-circular by induction on the length
of the context $\Gamma$.

\section{The definitions of types}

We will now define the types of a simple computational type theory without
universes. In the course of doing so, opportunities will arise for further
clarifying the position of the judgements, meaning explanations and proofs on
the one hand, and the propositions, definitions and verifications on the other
hand.

\subsection{The unit type}

First, we introduce two canonical forms with trivial reduction rules:
\begin{equation}
  \reduce\tyunit\tyunit\qquad
  \reduce\bullet\bullet
  \tag{Canonical}
\end{equation}

Next, we intend to make the judgement $\isset\tyunit$\ evident; and this is done by
defining what counts as a canonical verification of $\tyunit$\ and when two such
verifications are equal. To this end, we say that $\bullet$ is a canonical
verification of $\tyunit$, and that it is equal to itself.  I wish to emphasize
that this is the entire definition of the type: we have introduced syntax, and
we have defined the canonical forms, and there is nothing more to be done.

In the presentations of type theory which are currently in vogue, a type is
``defined'' by writing out a bunch of inference rules, but in type theory, the
definitions that we have given above are prior to the rules, which are
justified in respect of the definitions and the meaning explanations of the
judgements. For instance, based on the meaning of the various forms of sequent
judgement, the following rule schemes are justified:
\begin{gather*}
  \infer{
    \sequent\Gamma\isset\tyunit
  }{
  }\qquad
  \infer{
    \sequent\Gamma\eisset{\tyunit}{\tyunit}
  }{
  }\\[6pt]
  \infer{
    \sequent\Gamma\ver\bullet\tyunit
  }{
  }\qquad
  \infer{
    \sequent\Gamma\ever{\bullet}{\bullet}{\tyunit}
  }{
  }
\end{gather*}

Each of the assertions above has evidence of a certain kind; since the
justification of these rules with respect to the definitions of the logical
constants and the meaning explanations of the judgements is largely
self-evident, we omit it in nearly all cases. It is just important to remember
that it is not the rules which define the types; a type $A$ is defined in the
course of causing the judgement $\isset{A}$ to become evident.  These rules
merely codify standard patterns of use, nothing more, and they must each be
justified.

\subsection{The empty type}

The empty type is similarly easy to define. First, we introduce a constant:
\begin{gather*}
  \reduce\tyvoid\tyvoid\tag{Canonical}
\end{gather*}

To make the judgement $\isset\tyvoid$\ evident, we will say that there are no
canonical verifications of \tyvoid, and be done with it. This definition
validates some further rules schemes:
\begin{gather*}
  \infer{
    \sequent\Gamma\isset\tyvoid
  }{
  }\qquad
  \infer{
    \sequent\Gamma\eisset{\tyvoid}{\tyvoid}
  }{
  }\\[6pt]
  \infer{
    \sequent\Gamma\mathcal{J}
  }{
    \sequent\Gamma\ver{M}{\tyvoid}
  }
\end{gather*}

The last rule simply says that if we have a verification of $\tyvoid$, then we
may conclude any judgement whatsoever. Remember that the inference rules are
just notation for an \emph{evident} hypothetical judgement, e.g.\
$\hyp{\sequent{\Gamma}\mathcal{J}}{\sequent\Gamma\ver{M}{\tyvoid}}$.

Note that we did not introduce any special constant into the
computation system to represent the elimination of a verification of
$\tyvoid$\ (in proof-theoretic type theories, this non-canonical form
is usually called $\mathsf{abort}(R)$).  This is because,
computationally speaking, there is never any chance that we should
ever have use for such a term, since we need only consider the
evaluation of closed terms (which is guaranteed by the meaning
explanations), and by its very definition, there can never be a closed
verification of \tyvoid.

\subsection{The cartesian product of a family of types}

This will be our first dependent type, and it will likewise be the first
example of a non-trivial addition to the computation system. First, let us add
our canonical and non-canonical forms and their reduction rules:
\begin{gather*}
  \reduce{\typrod{A}{x}{B}}{\typrod{A}{x}{B}}
  \qquad
  \reduce{\lam{x}{E}}{\lam{x}{E}}
  \tag{Canonical}\\
  \infer{
    \reduce{\ap{M}{N}}{M'}
  }{
    \reduce{M}{\lam{x}{E}} &
    \reduce{[N/x]E}{M'}
  }\tag{Non-canonical}
\end{gather*}

This is the first time in this chapter that we have introduced a term former
with binding structure; it should be noted that the intensional equality of
expressions is up to alpha equivalence, and we will not pay attention to issues
of variable renaming in our presentation.

We will make evident the following judgement scheme:
\[
  \infer{
    \isset{\typrod{A}{x}{B}}
  }{
    \isset{A} &
    \sequent{x:A}{\isset{B}}
  }
\]

Or, written as a hypothetical judgement:
\[
  \hyp{\isset{\typrod{A}{x}{B}}}{\isset{A},\;\sequent{x:A}{\isset{B}}}
\]

This is to say, under the stated assumptions, we know what counts as a
canonical verification of $\typrod{A}{x}{B}$ and when two such
verifications are equal. We will say that $\lam{x}{E}$ is a canonical
verification of $\typrod{A}{x}{B}$ just when we know
$\sequent{x:A}{\ver{E}{B}}$; moreover, that two verifications
$\lam{x}{E}$ and $\lam{y}{E'}$ are equal just when
$\sequent{\ver{z}{A}}{\ever{[z/x]E}{[z/y]E'}{[z/x]B}}$.

By the meaning of the sequent judgement, this is to say that a lambda
expression must be functional with respect to its domain (i.e. it must take
equals to equals). We did not need to hypothesize directly two elements of the
domain and their equality because this is part of the meaning explanation for
the sequent judgement already. Likewise, two lambda expressions are equal when
equal inputs yield equal results in both.

The familiar inference rules, which codify the standard mode of use for the
family cartesian product, are justified by this definition:
\begin{gather*}
  \infer{
    \sequent\Gamma\isset{\typrod{A}{x}{B}}
  }{
    \sequent\Gamma\isset{A} &
    \sequent{\Gamma,x:A}{\isset{B}}
  }\qquad
  \infer{
    \sequent\Gamma\eisset{\typrod{A}{x}{B}}{\typrod{A'}{y}{B'}}
  }{
    \sequent\Gamma\eisset{A}{A'} &
    \sequent{\Gamma,z:A}{\eisset{[z/x]B}{[z/y]B'}}
  }\\[6pt]
  \infer{
    \sequent\Gamma\ver{\lam{x}{E}}{\typrod{A}{x}{B}}
  }{
    \sequent{\Gamma,x:A}{\ver{E}{B}}
  }\qquad
  \infer{
    \sequent\Gamma\ever{\lam{x}{E}}{\lam{y}{E'}}{\typrod{A}{x}{B}}
  }{
    \sequent{\Gamma,z:A}{\ever{[z/x]E}{[z/y]E'}{[z/x]B}}
  }\\[6pt]
  \infer{
    \sequent\Gamma\ver{\ap{M}{N}}{[N/x]B}
  }{
    \sequent\Gamma\ver{M}{\typrod{A}{x}{B}} &
    \sequent\Gamma\ver{N}{A}
  }\qquad
  \infer{
    \sequent\Gamma\ever{\ap{M}{N}}{\ap{M'}{N'}}{[N/x]B}
  }{
    \sequent\Gamma\ever{M}{M'}{\typrod{A}{x}{B}} &
    \sequent\Gamma\ever{N}{N'}{A'}
  }
\end{gather*}

Note that the type equality rule scheme that we gave above is structural; it is
validated by the meaning explanations, but it is by no means the full totality
of possible equalities between family cartesian product types, which is
extensional in this theory.

It will be instructive to explicitly justify the application rule above with
respect to the meaning explanations, since I have claimed that such rules are
posterior to the definitions we expounded prior to giving these rules.

\begin{proof}
It suffices to consider the case that $\reduce{\Gamma}{\cdot}$,
because hypotheses may always be added to the context (this is called
weakening). And so, by the meaning of the sequent judgement at the
empty context, the rule amounts to the assertion
\[
  \hyp{
    \ver{\ap{M}{N}}{[N/x]B}
  }{
    \ver{M}{\typrod{A}{x}{B}},\; \ver{N}{A}
  }
\]

By the meaning explanation for hypothetical judgement, and the definition of
the family cartesian product type, we know that $\reduce{M}{\lam{x}{E}}$ for
some $E$ such that we know $\sequent{x:A}{\ver{E}{B}}$; from the meaning of the
sequent judgement and the notion of an environment for a context, we can
conclude $\genj{L}{\hyp{\ver{[L/x]E}{[L/x]B}}{L\in A}}$. On the other hand,
from the computation rules, we know that if for some particular $E'$,
$\reduce{[N/x]E}{E'}$, then we know $\reduce{\ap{\lam{x}{E}}{N}}{E'}$; to
demonstrate the evidence of the premise, we may instantiate $L$ at $N$ to know
$\ver{[N/x]E}{[N/x]B}$, whence by the meaning of membership, we know that there
exists some canonical $E'$ such that $\reduce{[N/x]}{E'}$.

\end{proof}

\onehalfspacing

\chapter{Proof Theories and Logical Frameworks}

\section{Proof-theoretic type theory}

\subsection{Analytic and synthetic judgement}

A synthetic judgement is one for which the experience of \emph{coming
to know it} necessarily entails some knowledge which is not implicit
in the statement of the judgement; on the other hand, to know an
\emph{analytic} judgement is to know it purely on the basis of the
information contained inside it. So analytic judgements are decidable,
since if they may become evident, it will be purely on the basis of
their own content; whereas synthetic judgements become evident to
someone when they have obtained some particular evidence for them.

A logical theory has, then, both analytic and synthetic judgements;
the judgement $\isprop{P}$ is analytic, since its evidence follows
from the definition of $P$, whereas the assertion of $\istrue{P}$
entails the knowledge of some extra information, namely a verification
of $P$. When we have extended the logical theory to a type theory in
the manner of the previous chapter, the judgement $\ver{M}{P}$ is
also synthetic, since $\ver{M}{P}$ is not self-evident in general.

But why is it not enough to assert that $M$ verifies $P$ to know
whether $\ver{M}{P}$? It suffices to define a $P$ such that one cannot
decide in general whether some term is a verification of it. Let us
define the propositional symbol $\mathsf{P}$, and we intend to know
the judgement $\isprop{\mathsf{P}}$, whose meaning is to be expanded as
follows:
\begin{quote}
  To know $\isprop{\mathsf{P}}$ is to know counts as a canonical verification of $\mathsf{P}$.
\end{quote}

We will say, then, that $\bullet$ is a canonical verification of
$\mathsf{P}$ just when Goldbach's conjecture is true. Then it comes
immediately that the judgement $\ver{M}{\mathsf{P}}$ may not be known
or refuted on its own basis, nor even the judgement
$\ver{\bullet}{\mathsf{P}}$, since they depend on a proposition whose
truth is not known:

\begin{itemize}
  \item[] To know $\ver{M}{\mathsf{P}}$ is to know that $\reduce{M}{M'}$ to a
    canonical verification of $P$.
  \item[$\leadsto$] To know $\ver{M}{\mathsf{P}}$ is to know that
    $\reduce{M}{\bullet}$ such that Goldbach's conjecture is true.
\end{itemize}

\subsection{Proof of a judgement vs.\ verification of a proposition}

Because the judgement $\ver{M}{P}$ is synthetic, we cannot say that it
gives rise to a proof theory for the logic, since the core judgement
of a proof theory $\ofty{M}{A}$ must be analytic, in order to avoid the
infinite regress of a proof theory requiring a proof theory requiring
a proof theory, and so on.

The notion of verification of a proposition could never be the same as proof
anyway, except in the most trivial circumstances, since a verification is meant
to be an effective operation which realizes the truth of a proposition, and no
constraints whatsoever (termination, totality, etc.) are placed on these
operations except those which come from the meaning of the judgements (see
\cite{Dummett:Elements}, \cite{Prawitz12}, \cite{TroelstraA:conm}).

So a proof theory is necessarily intensional, and its judgements are to be
analytic/decidable. What is it, then, that we have considered so far which
corresponds with a proof $M$ such that $\ofty{M}{P}$ in a proof theory? As discussed
above, $M$ is not merely a term such that $\ver{M}{P}$, since this is not in
general enough information to know whether $M$ is a proof. In fact, $M$ must
comprise all the logical inferences which led to the knowledge that $P$ is
true, and so a meaning explanation for the judgement $\ofty{M}{P}$ in a proof theory
immediately suggests itself:
\begin{quote}
  To know $\ofty{M}{P}$ is to know that $M$ is evidence (demonstration, proof,
  derivation) of the judgement $\istrue{P}$.
\end{quote}

And so the term domain of the proof theory is not the same as the one that we
have considered so far; it must consist in terms which represent traces of the
inferences made in the course of knowing the judgements of a logical theory.
There is a sense in which one can consider the types of a proof theory to
interpret the judgements of the logical theory, and this methodology is called
``judgements as types'' (and this implies ``derivations as terms'').

What I am calling a ``proof-theoretic type theory'' is a type theory of the
sort used in the proof assistants Agda, Coq and Idris, whereas the kind of
type theory that I have described in the previous sections, the one based on
meaning explanations, underlies the proof assistant Nuprl.

The proof-theoretic type theories on the one hand are often called
``intensional'' and the computational type theories on the other hand are
usually ``extensional''; these characterizations are certainly true, though
they are not \emph{essential}; moreover, I fear that comparing one of the
former with one of the latter is not quite fair, since there is not any clear
analogy to be had. That is to say, the judgement $\ver{M}{P}$ is a judgement
which is added to a logical theory and its meaning is (briefly) ``$M$ evaluates
to a canonical verification of $P$'', whereas $\ofty{M}{P}$ cannot be construed
as a judgement added to a logical theory. Instead, it must be understood as
part of a (proof) theory which is overlayed atop an existing logical theory; it
is possible to understand the theory which contains the judgement $\ofty{M}{P}$
to be a metatheory, or logical framework, for the theory which contains the
judgement $\istrue{P}$, which can be construed as the ``object language''.

In short, the judgements $\ver{M}{P}$ and $\ofty{M}{P}$ are unrelated to each other in
two respects: firstly, that they have different meanings, and secondly that the
one is at the same level as the judgements of a logical theory, whereas the
latter is a judgement in a theory which is defined over a logical theory.

\section{Martin-L\"of's equational logical framework}

To make this more concrete, let us expound a proof theoretic type
theory called \MLLF, which stands for ``Martin-L\"of's (equational) logical
framework'';\footnote{For a detailed overview of Martin-L\"of's equational logical framework, see \cite{PiMLTT}.} in the course of introducing each type, we will specify
which judgement of the underlying logical theory it is meant to
interpret.

We start with four categorical judgements:\\

\begin{tabular}{c|l}
Judgement Form & Pronunciation \\ \hline
  $\type{\alpha}$ & $\alpha$ is a type \\
  $\etype{\alpha}{\beta}$ & $\alpha$ and $\beta$ are equal types \\
  $\ofty{M}{\alpha}$ & $M$ is of type $\alpha$ \\
  $\eofty{M}{N}{\alpha}$ & $M$ and $N$ are equal at type $\alpha$ \\
\end{tabular}\\

But we have not defined the meaning of the judgements; let us do so below:

\begin{quote}
  To know $\type\alpha$ is to know what counts as an object of type $\alpha$,
  and when two such objects are equal.
\end{quote}

For now, we'll leave the question of what is an ``object'' as
abstract; in many cases, types will represent judgements of a logical
theory, and the objects will be the derivations (demonstrations,
proofs) of those judgements.

\begin{quote}
  To know $\etype{\alpha}{\beta}$ is to know that any object of type $\alpha$ is
  also an object of type $\beta$, and two equal objects of type $\alpha$ are
  equal as objects of type $\beta$ (necessarily presupposing \type\alpha\ and
  \type\beta).
\end{quote}

\begin{quote}
  To know $\ofty{M}{\alpha}$ is to know that $M$ is an object of type $\alpha$
  (necessarily presupposing $\type\alpha$).
\end{quote}

\begin{quote}
  To know $\eofty{M}{N}{\alpha}$ is to know that $M$ and $N$ are equal objects of type
  $\alpha$ (necessarily presupposing $\ofty{M}{\alpha}$ and $\ofty{N}{\alpha}$).
\end{quote}

In addition to the above judgements, we will need contexts (with their
wellformedness judgement $\isctx{\Gamma}$) and an intensional sequent judgement
\framebox{$\lfsequent{\Gamma}{\mathcal{J}}$}; their meanings here will differ
from the sequent judgements of computational type theory, in that they must
mean proof-theoretic derivability, rather than semantic consequence.

At this point, we may begin adding types to the logical framework. In
practice, most types which we will introduce in the logical framework
will be defined in terms of a judgement of the logical theory which
lies below it.  For instance, hypothetical judgement in the logical
theory is represented by a function type in the logical framework,
$(x:\alpha)\beta$, whose typehood is meant to be evident under the
following circumstances
\[
  \infer{
    \type{(x:\alpha)\beta}
  }{
    \type\alpha &
    \lfsequent{x:\alpha}{\type\beta}
  }
\]
Or as a hypothetical judgement,
$\hyp{\type{(x:\alpha)\beta}}{\type\alpha,\lfsequent{x:\alpha}{\type\beta}}$.

Now, to know this judgement is to know that under the circumstances we
know what is an object of type $\alpha$ and when two such objects are
equal, and that if we have such an object $x$ of type $\alpha$, we
know what an object of type $\beta$ is, and when two such objects are
equal---then we know what an object of type $(x:\alpha)\beta$ is, and
moreover, for any two objects $y,z$ of type $\alpha$, that
$[y/x]\beta$ and $[z/x]\beta$ are equal as types. To make this
evident, then, we will say that under those circumstances an object of
type $(x:\alpha)\beta$ is an object $[x]M$ such that one knows
$\lfsequent{x:\alpha}{\ofty{M}{\beta}}$ and $\genj{x,y}{\hyp{\eofty{[y/x]M}{[z/x]M}{[y/x]\beta}}{\eofty{y}{z}{\alpha}}}$;
furthermore, two such objects are equal just when they yield equal
outputs for equal inputs.

Then, for each atomic proposition $P$, we can easily
define a type $\prf{P}$, as follows. Under the circumstances that
$\isprop{P}$ in the logical theory, then $\type{\prf{P}}$ in the
logical framework, since we will define an object of type $\prf{P}$ to
be a derivation of $\istrue{P}$; beyond reflexivity,
further definitional equalities can be added to reflect the harmony of
introduction and elimination rules.

Now, the definitions we have given for the types above are
``intuitively'' correct, but they actually fail to satisfy the meaning
explanation that we have given for $\type\alpha$, because they do not
take into account neutral terms. In the following sections, we will
investigate this problem in more detail and propose a solution.

\subsection{What is an ``object''?}
It is time to revisit what it means to be an ``object'' of a type in the
proof-theoretic type theory; we must note how this will necessarily differ from
what it meant to be a ``verification'' of a proposition in the previous
sections. Namely, a verification of a proposition is either a \emph{canonical
verification} of that proposition (and what sort of thing this might be is
known from the presupposition $\isprop{P}$), or it is a means of getting such
a canonical verification (i.e. a term which evaluates to a canonical
verification).

On the other hand, what we have called an ``object'' of type $P$ is quite
different, since in addition to the possibility that it is a canonical proof of
the judgement $\istrue{P}$, it may also be \emph{neutral} (i.e. blocked by a
variable); we will call this ``normal'' rather than ``canonical''. Why does
this happen?

In order to keep the judgement $\ofty{M}{A}$ analytic (decidable), its meaning
explanation can no longer be based on the idea of the computation of closed
terms to canonical form; instead, we will consider the computation of open
terms (i.e. terms with free variables) to \emph{normal} form. The desire for
$\ofty{M}{A}$ to be analytic follows from our intention that it characterize a
\emph{proof theory}: we must be able to recognize a proof when we see one. But
why are closed-term-based meaning explanations incompatible with this goal?
Consider briefly the following judgement:
\[
  \genj{n}{\hyp{\ver{M(n)}{P}}{\ver{n}{\naturals}}}
\]

To know this judgement is to know that $M(n)$ computes to a canonical
verification of $P$ whenever $n$ is a natural number; when $P$'s use of $n$ is
not trivial, this amounts to testing an infinite domain (all of the natural
numbers), probably by means of mathematical induction. The judgement is then
clearly synthetic: to know it is, briefly, to have come up with an (inductive)
argument that $M(N)$ computes to a canonical verification of $P$ at each
natural number $n$.

On the other hand, the judgement $\lfsequent{n:\naturals}{\ofty{M(n)}{P}}$ must have a
different meaning, one which admits its evidence or refutation purely on
syntactic/analytic grounds. In essence, it is to know that $M(n)$ is a proof of
$P$ for any \emph{arbitrary} object/expression $n$ such that $\ofty{n}{\naturals}$
(i.e., the only thing we know about $n$ is that it is of type $\naturals$; we
do not necessarily know that it is a numeral).

\section{A critique of \textbf{MLLF}}

The type theory which we constructed in the previous section is to be
considered a proof theory for a logic with the judgements
$\isprop{P}$, $\istrue{P}$ and
$\hyp{\mathcal{J}}{\mathcal{J}'}$. There are a few reasons to be
dissatisfied with this state of affairs, which I shall enumerate in
this section.

\subsection{Lack of computational content}

Unlike the type theory in the first chapter, there is no built-in computational
content. In a computational type theory which is defined by the verificationist
meaning explanations, the computational content of terms is understood
immediately by means of the $\reduce{M}{M'}$ relation; that is, computation is
prior to the main judgements because their meaning explanations are defined in
terms of evaluation to canonical form.

On the other hand, in the type theory above we did not give a
primitive reduction relation; instead, we simply permitted the
endowement of proofs with definitional equalities which reflect the
harmony of introduction and elimination rules. That is, if we have
known the judgement $\istrue{P}$ by means of an indirect argument
(derivation), it must be the case that this derivation corresponds to
a direct one; we reflect this in the proof theory by defining the
indirect derivation to be definitionally equal to the direct one.

However, this does not amount to computational content being present in terms:
only \emph{post facto} may the definitional equality be construed as giving
rise to computation, through a metamathematical argument which shows that the
definitional equality is confluent and can be used to define a functional
normalization relation.

And this is the reason for the peculiarity of the proof-theoretic
meaning explanations, namely that they do not include phrases like
``evaluates to a canonical...'', since evaluation may only be
understood after taking the meanings of the judgements ($\type\alpha$,
$\etype{\alpha}{\beta}$, $\ofty{M}{\alpha}$, $\eofty{M}{N}{\alpha}$) as giving rise to a
closed formal system which is susceptible to metamathematical
argument: to refer to evaluation in the meaning explanations for the
core judgements, then, would be impredicative.

\subsection{Modularity of definition}

By the same token, the distinction between canonical (direct) and non-canonical
(indirect) proof may not be understood as a core notion in the theory, but must
be understood separately, secondarily. Why is this a problem? It means that the
definition of each type must be made with the full knowledge of the definitions
of every other type; in essence, the open-ended nature of type theory is
obliterated and one is forced into a fixed formal system; this is in addition
to the fact that it causes the epistemic content of $\type\alpha$ for any type
$\alpha$ to be extremely complicated.

To illustrate, let us consider as an example a type theory which has
four type-formers: trivial truth $\top$, trivial falsity $\bot$,
implication $(\alpha)\beta$, and conjunction $\product\alpha\beta$; we
will then introduce the following terms to represent proofs: the
trivial element $\bullet$, \emph{reductio ad absurdum}
$\abort{\alpha}{E}$, abstraction $[x:\alpha]E$, application $E(E')$,
pairing $\pair{E}{E'}$, and projections $\fst{E}$, $\snd{E}$.

If we will try to make the judgement $\type\top$ evident, the deficiencies of
the formulation will immediately present themselves.

\begin{quote}
To know $\type\top$ is to know what counts as an object of type $\top$, and when
two such objects are equal. An object of type $\top$, then, is either the
expression $\bullet$, or an expression $\abort{\top}{E}$ such that we know
$\ofty{E}{\bot}$, or an expression $E(E')$ such that we know $\ofty{E}{(\alpha)\top}$ and
$\ofty{E'}{\alpha}$, or an expression $\fst{E}$ such that we know $\ofty{E}{\product\top\beta}$
for some $\beta$, or an expression $\snd{E}$ such that we know
$\ofty{E}{\product\alpha\top}$ for some $\alpha$; and we additionally have that $\bullet$
is equal to $\bullet$, and ...
\end{quote}

To save space, we elide the rest of the definition of equality for $\top$; what
we have seen so far already suffices to bring to light a serious problem: the
definition of any type requires knowledge of the entire syntax of the theory.
The judgement $\type\alpha$ may never be made evident in isolation, but must be
done with full understanding of all the other types and their definitions.

Furthermore, to extend an existing theory with a new type, the definitions of
every other type must be rewritten to account for the elimination forms of the
new type.

\section{A way forward: verifications \& uses}

The second critique of \MLLF{} may be partially addressed by fragmenting type
theory into a logic of \emph{verifications \& uses}: instead of a type being
defined by its introduction rules, it must be simultaneously defined by its
introduction rules (verifications) and its elimination rules (uses). In
practice, this amounts to a standard technique known as \emph{bidirectional
type checking}.

The semantic priority of the forms of judgement also changes drastically: the
sequent judgement must in this case be explained \emph{before} the categorical
judgements; moreover, sequents may no longer be explained modularly in terms of
general and hypothetical judgement, since the latter amounts to
\emph{semantic} consequence (admissibility), whereas the meaning of a sequent
in a proof theory should be \emph{syntactic} consequence (derivability).

Because the target theory lacks computation, it is necessary to rule out
redexes from terms syntactically, but this complicates the definition of
substitution; to address this, Watkins introduced in \cite{watkins} a technique
known as \emph{hereditary substitution}, which is a family of syntax-directed
(algorithmic) judgements which contract redexes along the way, guaranteeing
canonical form in their outputs. Both bidirectional type checking and
hereditary substitutions have been used to great effect in the descendants of
the Edinburgh Logical Framework (see \cite{Harper-Licata-2007}).

The first critique, the lack of computational content, is more difficult to
address. Roughly, the right way to do it is to replace the notion of the
evaluation of closed terms to canonical form with the evaluation of \emph{open}
terms to normal form. Peter Dybjer demonstrates in \cite{series/leus/Dybjer12a}
how this technique may be used to endow the Calculus of Constructions with a
meaning explanation, albeit necessarily of a very different kind than we have
considered here.

\bibliographystyle{plain}
\nocite{*}
\bibliography{refs}

\end{document}